\documentclass[a4paper,11pt]{article}
\pdfoutput=1
\usepackage{jcappub}
\usepackage{mathtools}

\graphicspath{{plots/}}

\newcommand{\lhs}{\lambda_{hs}}
\newcommand{\mdm}{m_s}
\newcommand{\rR}{\rho_R}
\newcommand{\rp}{\rho_\phi}
\newcommand{\Gp}{\Gamma_\phi}
\newcommand{\gs}{g_\star}
\newcommand{\gss}{g_{\star s}}
\newcommand{\arh}{a_\text{rh}}
\newcommand{\Trh}{T_\text{rh}}
\newcommand{\Hrh}{H_\text{rh}}
\newcommand{\sv}{\langle\sigma v\rangle}

\title{From WIMPs to FIMPs with\\Low Reheating Temperatures}

\author[a]{Javier Silva-Malpartida} \emailAdd{javier.silvam@pucp.edu.pe}
\author[b]{Nicolás Bernal} \emailAdd{nicolas.bernal@nyu.edu}
\author[a]{\\Joel Jones-Pérez} \emailAdd{jones.j@pucp.edu.pe}
\author[c]{and Roberto A. Lineros} \emailAdd{roberto.lineros@ucn.cl}

\affiliation[a]{Sección Física, Departamento de Ciencias, Pontificia Universidad Católica del Perú. Apartado 1761, Lima, Peru.}
\affiliation[b]{New York University Abu Dhabi. PO Box 129188, Saadiyat Island, Abu Dhabi, United Arab Emirates.}
\affiliation[c]{Departamento de Física, Universidad Católica del Norte. Avenida Angamos 0610, Casilla 1280, Antofagasta, Chile.}

\abstract{
Weakly- and Feebly-Interacting Massive Particles (WIMPs and FIMPs) are among the best-motivated dark matter (DM) candidates. In this paper, we investigate the production of DM through the WIMP and FIMP mechanisms during inflationary reheating. We show that the details of the reheating, such as the inflaton potential and the reheating temperature, have a strong impact on the genesis of DM. The strong entropy injection caused by the inflaton decay has to be compensated by a reduction of the portal coupling in the case of WIMPs, or by an increase in the case of FIMPs. We pinpoint the smooth transition between the WIMP and the FIMP regimes in the case of low reheating temperature. As an example, we perform a full numerical analysis of the singlet-scalar DM model; however, our results are generic and adaptable to other particle DM candidates.
Interestingly, in the singlet-scalar DM model with low-reheating temperature, regions favored by the FIMP mechanism are already being tested by direct detection experiments such as LZ and XENONnT.
}

\begin{document}
\maketitle

\section{Introduction}
\label{sec:intro}
The existence of nonbaryonic dark matter (DM) in the Universe is compelling, as suggested by astrophysical and cosmological observations~\cite{Bertone:2016nfn, Planck:2018vyg}.
The most commonly assumed production mechanism for DM in the early universe corresponds to the weakly interacting massive particle (WIMP) paradigm~\cite{Arcadi:2017kky, Roszkowski:2017nbc}, in which DM has mass and couplings at the electroweak scale. Here, WIMPs reach chemical equilibrium with the standard model (SM) thermal plasma and eventually freeze out, giving rise to the observed DM relic abundance.
A thermally averaged annihilation cross-section $\sv \sim \mathcal{O}(10^{-26})$~cm$^3$/s is typically required to fit observations~\cite{Steigman:2012nb}.
At some point in the past, the community was particularly interested in the WIMP mechanism because it was expected to be tested in several complementary ways, including direct, indirect, and collider probes.
However, destiny is cruel: the current null experimental results and severe constraints on the natural parameter space are forcing us to search beyond the standard WIMP paradigm~\cite{Arcadi:2017kky}.

Relaxation of the condition of equilibrium reveals a straightforward alternative to the WIMP paradigm. Assuming instead that DM particles never entered thermal equilibrium with the SM plasma, the present abundance of DM may have been produced by the so-called freeze-in mechanism~\cite{McDonald:2001vt, Choi:2005vq, Kusenko:2006rh, Petraki:2007gq, Hall:2009bx, Elahi:2014fsa, Bernal:2017kxu}. Because of the feeble interaction strength that the mechanism requires, these DM candidates are usually called feebly-interacting massive particles (FIMPs). Depending on the moment at which they are produced, FIMPs could be infrared (IR) or ultraviolet (UV)~\cite{Elahi:2014fsa}. Although IR FIMPs are characteristic of renormalizable interactions with very suppressed couplings $\sim \mathcal{O}(10^{-11})$, UV FIMPs come from non-renormalizable interactions suppressed by a high dimensionful scale~\cite{Bernal:2017kxu}. Finally, in the presence of a thermal partner, FIMPs can also be produced by the late decays of the former~\cite{Feng:2003xh, Feng:2003uy, Asaka:2005cn, Garny:2018ali, Faber:2019mti}.

It is important to note that for estimating the DM relic abundance, one has to have under control not only particle physics but also cosmology.
A common minimalistic assumption is the so-called standard cosmology, in which the energy density of the universe was dominated by SM radiation from the end of inflationary reheating until the onset of Big Bang nucleosynthesis (BBN) at $t \sim 1$~s.
Additionally, the end of reheating is assumed to occur at a very high temperature $\Trh$, the latter characterizing the beginning of the radiation-dominated era.
However, there are no indispensable reasons to assume so~\cite{Allahverdi:2020bys}.
DM production with a low reheating temperature has been intensively studied in the literature, usually triggered by the decay of a long-lived massive particle~\cite{Giudice:2000ex, Fornengo:2002db, Pallis:2004yy, Gelmini:2006pw, Drees:2006vh, Yaguna:2011ei, Roszkowski:2014lga, Drees:2017iod, Bernal:2018ins, Bernal:2018kcw, Cosme:2020mck, Arias:2021rer, Bernal:2022wck, Bhattiprolu:2022sdd, Haque:2023yra, Ghosh:2023tyz}, or by Hawking evaporation of primordial black holes~\cite{Green:1999yh, Khlopov:2004tn, Dai:2009hx, Fujita:2014hha, Allahverdi:2017sks, Lennon:2017tqq, Morrison:2018xla, Hooper:2019gtx, Chaudhuri:2020wjo, Masina:2020xhk, Baldes:2020nuv, Gondolo:2020uqv, Bernal:2020kse, Bernal:2020ili, Bernal:2020bjf, Bernal:2021akf, Cheek:2021odj, Cheek:2021cfe, Bernal:2021yyb, Bernal:2021bbv, Bernal:2022oha, Cheek:2022dbx, Mazde:2022sdx, Cheek:2022mmy}.\footnote{For studies on baryogenesis with a low reheating temperature or during an early matter-dominated phase, see Refs.~\cite{Davidson:2000dw, Giudice:2000ex, Allahverdi:2010im, Beniwal:2017eik, Allahverdi:2017edd} and~\cite{Bernal:2017zvx, Chen:2019etb, Bernal:2022pue, Chakraborty:2022gob}, respectively. Furthermore, the production of primordial gravitational waves in scenarios with an early matter era has recently received particular attention~\cite{Assadullahi:2009nf, Durrer:2011bi, Alabidi:2013lya, DEramo:2019tit, Bernal:2019lpc, Figueroa:2019paj, Bernal:2020ywq}.}
Finally, we note that if DM does not reach equilibrium with the visible sector, the possible overproduction of DM during inflation could be relaxed in scenarios with low reheating temperature~\cite{Bernal:2018hjm, Garcia:2022vwm, Kaneta:2022gug, Garcia:2023qab, Cosme:2023xpa}.

In this paper, we investigate the genesis of DM during reheating in scenarios with a low reheating temperature, where the inflaton field $\phi$ oscillates around the minimum of a generic monomial potential $V(\phi) \propto \phi^n$.
Such potentials naturally arise in several inflationary scenarios like the $\alpha$-attractor $T$- and $E$-model~\cite{Kallosh:2013hoa, Kallosh:2013maa, Kallosh:2013yoa}, or the Starobinsky model~\cite{Starobinsky:1980te, Starobinsky:1981vz, Starobinsky:1983zz, Kofman:1985aw}.
In this case, the equation of state (EoS) parameter and the decay rate of the inflaton are modified compared to the case usually assumed, where the inflaton oscillates in a quadratic potential (that is, $n = 2$). In particular, the EoS depends on the shape of the potential $\omega = (n-2)/(n+2)$~\cite{Turner:1983he}, and the decay width of the inflaton develops a time dependency~\cite{Garcia:2020eof, Garcia:2020wiy}. As a consequence, the inflaton and SM energy densities exhibit a non-standard evolution during reheating~\cite{Co:2020xaf}, which in turn has a strong impact on DM production.\footnote{We note that the phenomenology of FIMPs~\cite{Garcia:2020eof, Garcia:2020wiy, Ahmed:2021fvt, Barman:2022tzk, Ahmed:2022tfm, Becker:2023tvd} and the QCD axion DM~\cite{Arias:2022qjt} with a time-dependent decay rate has recently been analyzed.}

An important aspect of our work lies in the numerical evaluation of DM interaction rates. To this end, we have developed a code allowing the evaluation of DM interactions through \texttt{MadGraph}~\cite{Alwall:2014hca} for any model implemented in the \texttt{UFO} format~\cite{Degrande:2011ua}. Along with the numerical solving of the corresponding Boltzmann equations for the DM and the background, this allows us to carefully keep track of the DM evolution. In particular, we find that for low-reheating temperatures, there is a smooth transition between the WIMP and IR FIMP regimes. We illustrate our results through the singlet scalar DM (SSDM) model~\cite{Silveira:1985rk, McDonald:1993ex, Burgess:2000yq}.

The manuscript is organized as follows. In Section~\ref{sec:setup}, we first introduce the singlet scalar DM model and then explain the evolution of the inflaton and the SM energy densities during the reheating period. In Section~\ref{sec:DM_density}, we study the generation of the DM relic density for the WIMP and FIMP limits of the SSDM. In particular, we highlight the smooth transition between the WIMP and the FIMP regimes, in the case of a low-reheating scenario.
The conclusions are given in Section~\ref{sec:concl}.

\section{Setup} \label{sec:setup}
In this section, the setup of the analysis is presented; first the particle physics model, and then the cosmological model during reheating. 

\subsection{Singlet Scalar Dark Matter} 
The SSDM model~\cite{Silveira:1985rk, McDonald:1993ex, Burgess:2000yq} extends the SM by adding a real scalar field, $s$, which is a singlet under the SM gauge symmetry $SU(3)_c \otimes SU(2)_L \otimes U(1)_Y$. It includes a $\mathbb{Z}_2$ parity, under which only $s$ is odd. With these considerations, the most general renormalizable scalar potential of the model is
\begin{equation}
    V = \lambda_H\left(|H|^2-v_H^2\right)^2 + \mu_s^2\,s^2 + \lambda_s\,s^4 + \lhs\,|H|^2\,s^2,
\end{equation}
where $H$ is the SM Higgs doublet, and $v_H \equiv \mu_H/\sqrt{2\lambda_H}$ is the Higgs vacuum expectation value (vev). This potential contains only three new free parameters: $i)$ $\mu^2_s$ which contributes to the mass $\mdm$ of $s$, $ii)$ a quartic coupling $\lambda_s$, and $iii)$ the Higgs portal coupling $\lhs$, which is the only connection (besides gravity) between the dark and visible sectors.
Taking them all as positive, only $H$ acquires a vev, leading to the following mass term for $s$
\begin{equation}
    m^2_s = 2\,\mu_s^2+\frac{\lhs}{\lambda_H}\mu^2_H\,.
\end{equation}
Given that $s$ is the only odd particle under $\mathbb{Z}_2$, it is automatically stable and must appear in even numbers within the interaction vertices. It is also a viable candidate for the DM of the universe.

The phenomenology of the singlet scalar model as a candidate for DM has been widely studied in the literature, especially in the case where the Higgs portal is of order one $\lhs \sim \mathcal{O}(1)$, so that DM could have been thermally produced in the early universe.
In this case, the model can be probed at colliders~\cite{Barger:2007im, Kanemura:2011nm, Djouadi:2011aa, No:2013wsa, Craig:2014lda, Robens:2015gla, Han:2016gyy, Ruhdorfer:2019utl, Englert:2020gcp, Garcia-Abenza:2020xkk, Biekotter:2022ckj}, with direct and indirect DM searches~\cite{Yaguna:2008hd, Goudelis:2009zz, Profumo:2010kp, Urbano:2014hda, Duerr:2015mva, DiMauro:2023tho}, or in a combination of different experimental techniques~\cite{He:2009yd, Djouadi:2012zc, Damgaard:2013kva, Baek:2014jga, Curtin:2014jma, Feng:2014vea, Han:2015hda, Duerr:2015aka, Benito:2016kyp, GAMBIT:2018eea}. In particular, combined fits assuming standard cosmology~\cite{Cline:2013gha, Beniwal:2015sdl, GAMBIT:2017gge, Athron:2018ipf} indicate that $s$ can still be a good thermal DM candidate, but only in a very reduced region of its parameter space.
On the other hand, if the Higgs portal is much smaller $\lhs \sim \mathcal{O}(10^{-11})$, the scalar singlet DM could have been produced non-thermally~\cite{Yaguna:2011qn, Yaguna:2011ei, Campbell:2015fra, Kang:2015aqa, Belanger:2018ccd, Bernal:2018kcw, Heeba:2018wtf, Huo:2019bjf, Lebedev:2019ton, Cosme:2023xpa}.
Alternatively, if the DM quartic coupling is sizeable $\lambda_s \sim \mathcal{O}(1)$, the scalar singlet could have been produced thermally, but from self-production and cannibalization reactions~\cite{Carlson:1992fn, Bernal:2015xba, Heikinheimo:2017ofk, Arcadi:2019oxh, Bernal:2020gzm}.
We note that this model has also been studied in the framework of nonstandard cosmology~\cite{Yaguna:2011ei, Bernal:2018ins, Hardy:2018bph, Bernal:2018kcw, Allahverdi:2020bys}, and in the context of Hawking evaporation of primordial black holes~\cite{Bernal:2020bjf}.

To track the evolution of the DM density, we have developed a code that executes the automatic calculation of the thermally-averaged annihilation cross-section $\sv$, for any particle-physics model.
Our method is based on the \texttt{standalone} subroutine from \texttt{MadGraph}, which calculates all tree-level amplitudes contributing to $\sv$, as long as the appropriate \texttt{UFO} and \texttt{param\_card} files are provided. For the SSDM model, we have generated the \texttt{UFO} files using \texttt{SARAH}~\cite{Staub:2008uz, Staub:2012pb, Staub:2013tta}. The \texttt{param\_card} files were generated in turn by \texttt{SPheno}~\cite{Porod:2003um, Porod:2011nf, Staub:2011dp}. 

In the case of the SSDM, as the square matrix elements $\left | \mathcal{M} \right |^2$ do not depend on the solid angle, only on the center-of-mass energy $\sqrt{s}$, the integration over the phase space is trivial.
Then, we combined all the processes and obtained the cross section as a function of the center-of-mass energy, $\sigma (s)$. Finally, we compute the thermally-averaged DM annihilation cross-section using the standard relation~\cite{Gondolo:1990dk}
\begin{equation} \label{sigv}
    \sv(T) = \int_{4\mdm^2}^{\infty} \mathrm{d}s\, \frac{(s-4\mdm^2) \, \sqrt{s} \, K_{1}\left ( \sqrt{s}/T \right ) \sigma(s)}{8 \,T\, \mdm^{4}\, K_2^2\left ( \mdm/T \right ) }\,,
\end{equation}
where $K_i$ is the modified Bessel function. We perform a numerical integration of Eq.~\eqref{sigv} using the \texttt{adaptive Simpson's} method and compute $\left \langle \sigma \upsilon \right \rangle$ as a function of the temperature $T$ of the SM plasma.

\subsection{Cosmology During Reheating} \label{sec:reheating}
We consider an inflaton $\phi$ that, after the end of inflation, oscillates at the bottom of a monomial potential $V(\phi)$ of the form
\begin{equation}
    V(\phi) = \lambda_\phi\, \frac{\phi^n}{\Lambda^{n - 4}}\,,
\end{equation}
where $\lambda_\phi$ is a dimensionless coupling and $\Lambda$ an energy scale.
Potentials with these types of minimum naturally arise in several inflationary scenarios, such as the $\alpha$-attractor $T$- and $E$-model~\cite{Kallosh:2013hoa, Kallosh:2013maa, Kallosh:2013yoa}, or the Starobinsky model~\cite{Starobinsky:1980te, Starobinsky:1981vz, Starobinsky:1983zz, Kofman:1985aw}.
The evolution of the energy density $\rp$ of the homogeneous inflaton that coherently oscillates at the bottom of the potential $V(\phi)$ is given by~\cite{Turner:1983he, Garcia:2020wiy, Bernal:2022wck}
\begin{equation} \label{eq:drhodt}
    \frac{d\rp}{dt} + \frac{6\, n}{2 + n}\, H\, \rp = - \frac{2\, n}{2 + n}\, \Gp\, \rp\,,
\end{equation}
where $\Gp$ is the inflaton decay width, and $H$ the Hubble expansion rate.

Before proceeding with the full numerical study of the reheating, it is useful to have an analytical understanding of its dynamics~\cite{Bernal:2022wck}.
During reheating, that is when the scale factor $a$ is in the range $a_I \ll a \ll \arh$, with $a_I$ and $\arh$ corresponding to the scale factors at the beginning and end of reheating, respectively, Eq.~\eqref{eq:drhodt} admits the analytical solution
\begin{equation} \label{eq:rpsol}
    \rp(a) \simeq \rp (\arh) \left(\frac{\arh}{a}\right)^\frac{6\, n}{2 + n}.
\end{equation}
As in this epoch the Hubble expansion rate is dominated by the inflaton energy density, it follows that
\begin{equation} \label{eq:Hubble}
    H(a) \simeq \Hrh \times
    \begin{dcases}
        \left(\frac{\arh}{a}\right)^\frac{3\, n}{n + 2} &\text{ for } a \leq \arh\,,\\
        \left(\frac{\arh}{a}\right)^2 &\text{ for } \arh \leq a\,,
    \end{dcases}
\end{equation}
with $\Hrh \equiv H(\Trh)$.
At the end of reheating (that is, at $a = \arh$), the energy densities of radiation and inflaton are equal, $\rR(\arh) = \rp(\arh) = 3\, M_P^2\, \Hrh^2$.
Note that to avoid spoilage of BBN, the reheating temperature must satisfy $\Trh \geq T_\text{BBN} \simeq 4$~MeV~\cite{Sarkar:1995dd, Kawasaki:2000en, Hannestad:2004px, DeBernardis:2008zz, deSalas:2015glj}.

During reheating, the inflaton transfers its energy density to SM radiation, and the end of reheating corresponds to the onset of the SM radiation-dominated era.
The density of SM radiation energy $\rR$ is governed by the Boltzmann equation
\begin{equation} \label{eq:rR}
    \frac{d\rR}{dt} + 4\, H\, \rR = + \frac{2\, n}{2 + n}\, \Gp\, \rp\,.
\end{equation}
Using Eq.~\eqref{eq:rpsol}, one can solve Eq.~\eqref{eq:rR} and further obtain the solution for the evolution of the radiation energy density during reheating 
\begin{equation} \label{eq:rR_int}
    \rR(a) \simeq \frac{2\, \sqrt{3}\, n}{2 + n}\, \frac{M_P}{a^4} \int_{a_I}^a \Gp(a')\, \sqrt{\rp(a')}\, a'^3\, da',
\end{equation}
where a general scale factor dependence of $\Gp$ has been assumed.
Such a dependence can come, for example, from the inflaton mass parameter.
The effective mass $m_\phi$ for the inflaton field understood as the second derivative of its potential is given by
\begin{equation} \label{eq:mass}
    m_\phi^2 \simeq n\, (n-1)\, \lambda_\phi^{2/n}\, \Lambda^{2\, (4 - n)/n} \rp(a)^{(n-2)/n}. 
\end{equation}
It is interesting to note that for $n \neq 2$, $m_\phi$ features a field dependence that, in turn, leads to an inflaton decay rate with a scale factor (or time) dependence.

In the following, two different reheating scenarios in which the inflaton either decays to a pair of fermions or scalars will be investigated.
We assume that the inflaton only decays into a pair of fermions $\Psi$ and $\bar\Psi$ via a trilinear interaction $y\, \phi\, \bar\Psi\, \Psi$, with $y$ being the corresponding Yukawa coupling, or into a pair of scalars $S$ through the interaction $\mu\, \phi\, S\, S$, with $\mu$ being a dimension-full coupling.
The decay rate is therefore given by
\begin{equation} \label{eq:fer_gamma}
    \Gp =
    \begin{dcases}
        \frac{y_{\text{eff}}^2}{8\pi}\, m_\phi &\text{ for fermions,}\\
        \frac{\mu_{\text{eff}}^2}{8\pi\, m_\phi} &\text{ for scalars,}
    \end{dcases}
\end{equation}
where the effective couplings $y_\text{eff} \ne y$ and $\mu_\text{eff} \ne \mu$ (for $n \neq 2$) are obtained after averaging over oscillations~\cite{Shtanov:1994ce, Garcia:2020wiy, Ichikawa:2008ne}.
It is interesting to note the different mass dependence in the decay width, cf. Eq.~\eqref{eq:fer_gamma}, due to the different mass dimensions of the couplings.
By solving the SM energy density in Eq.~\eqref{eq:rR_int} and using the fact that it is related to the temperature $T$ of the SM bath by
\begin{equation}
    \rR(T) \equiv \frac{\pi^2}{30}\, \gs(T)\, T^4,
\end{equation}
where $\gs$ counts the number of relativistic degrees of freedom contributing to $\rR$~\cite{Drees:2015exa}, one can find that during reheating $T$ evolves as
\begin{equation} \label{eq:Tfer}
    T(a) \simeq \Trh \times
    \begin{dcases}
        \left(\frac{\arh}{a}\right)^{\frac32 \frac{n - 1}{n + 2}} &\text{ for fermions},\\
        \left(\frac{\arh}{a}\right)^{\frac32 \frac{1}{2 + n}} &\text{ for scalars},
    \end{dcases}
    \qquad(a<\arh)
\end{equation}
where $\Trh \equiv T(\arh)$ is the SM bath temperature at the end of reheating.
The Hubble expansion rate (cf. Eq.~\eqref{eq:Hubble}) can then be rewritten as a function of $T$ as
\begin{equation} \label{eq:Hfer}
    H(T) \simeq \Hrh \times
    \begin{dcases}
        \left(\frac{T}{\Trh}\right)^\frac{2\, n}{n - 1} &\text{ for fermions},\\
        \left(\frac{T}{\Trh}\right)^{2\, n} &\text{ for scalars}.
    \end{dcases}
    \qquad(T>\Trh)
\end{equation}
It is important to note that the fermionic case in Eqs.~\eqref{eq:Tfer} and~\eqref{eq:Hfer} is only valid if $n < 7$.\footnote{For $n\geq7$, the solution to Eq.~\eqref{eq:rR_int} leads to a different scaling for $T$ and $H$, see Ref.~\cite{Bernal:2022wck}. In particular, for $n>7$, the scaling of $T$ resembles the one for the standard cosmological case, i.e. $T(a) \propto a^{-1}$.}
Before closing this subsection, we note that for the case with $n = 2$ where the inflaton oscillates in a quadratic potential with an EoS $\omega = 0$, the standard dependence with the scale factor $\rR(a) \propto a^{-3/2}$ and $T(a) \propto a^{-3/8}$ is recovered.\\

After the end of reheating, that is, for $a \geq \arh$ or equivalently for $T \leq \Trh$, the SM entropy is conserved and the SM temperature follows the standard evolution given by
\begin{equation} \label{eq:TSM}
    T(a) = \left(\frac{\gss(\Trh)}{\gss(T)}\right)^\frac13 \Trh\, \frac{\arh}{a}\,,
    \qquad(a>\arh)
\end{equation}
where $\gss$ corresponds to the effective number of relativistic degrees of freedom contributing to the SM entropy~\cite{Drees:2015exa}, and the Hubble expansion rate
\begin{equation}
    H(T) \simeq \sqrt{\frac{\rR(T)}{3 M_P^2}} = \frac{\pi}{3} \sqrt{\frac{\gss(T)}{10}} \frac{T^2}{M_P}\,.
    \qquad(T<\Trh)
\end{equation}\\

Now that the dynamics of reheating has been analytically understood, a full numerical study can be carried out. To this end, using a fourth-order \texttt{Runge-Kutta} method we numerically solve the system of Boltzmann equations~\eqref{eq:drhodt} and~\eqref{eq:rR}, together with the inflaton mass  in Eq.~\eqref{eq:mass}, the decay rate in Eq.~\eqref{eq:fer_gamma}, and the full Hubble rate
\begin{equation} \label{eq:H}
    H = \sqrt{\frac{\rR + \rp}{3 M_P^2}}\,.
\end{equation}
For the initial conditions, we take $\rR = 0$ and tune $\rp$ so that we have a reheating (that is, the equality $\rR = \rp$) at the desired temperature $\Trh$.
Here, we focus on a scenario in which the inflaton exclusively decays into scalars or fermions, with $n = 2$ or 4. 

\begin{figure}[t]
    \def\sepf{0.495}
    \centering
    \includegraphics[width=\sepf\columnwidth]{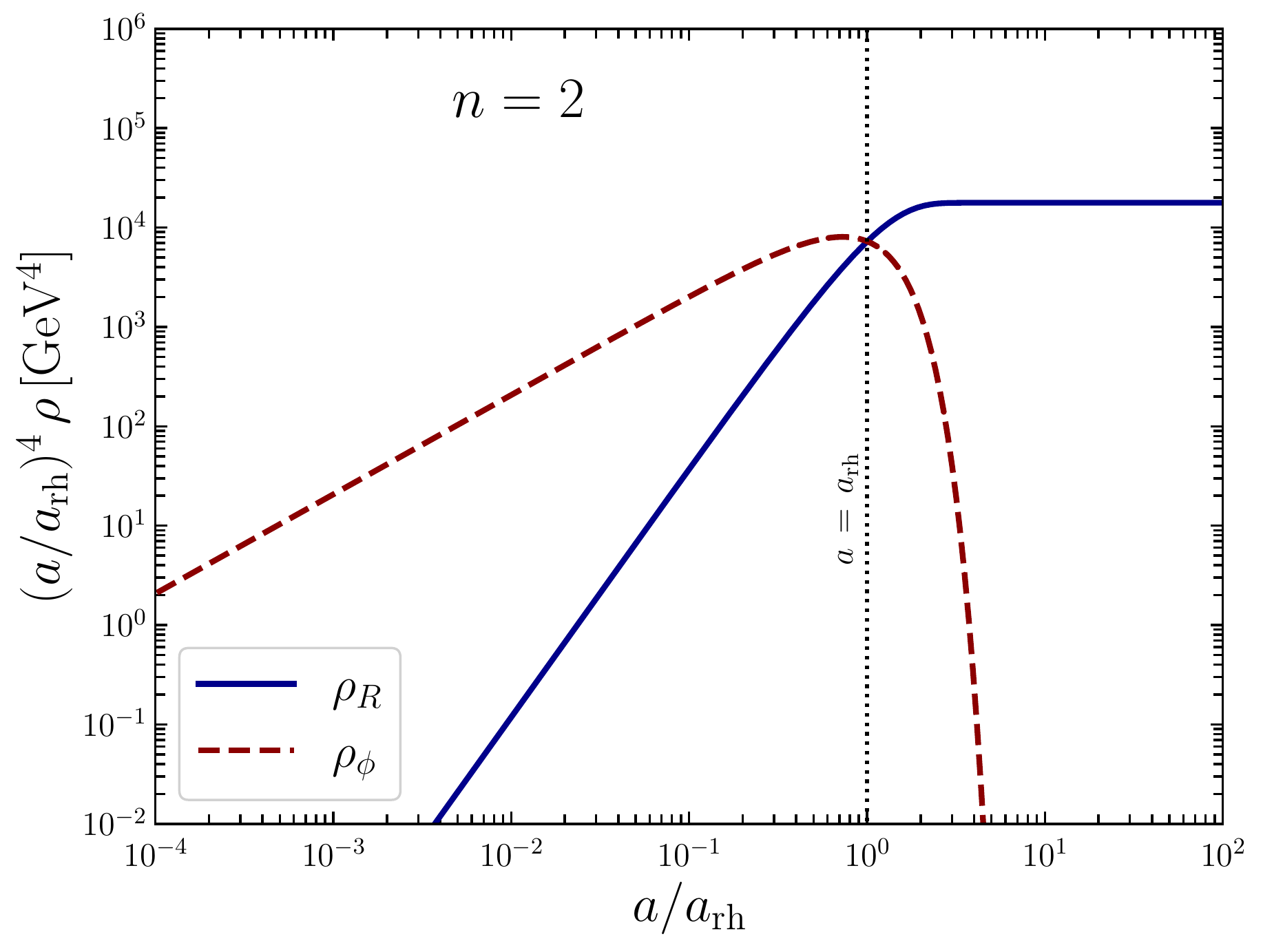}
    \includegraphics[width=\sepf\columnwidth]{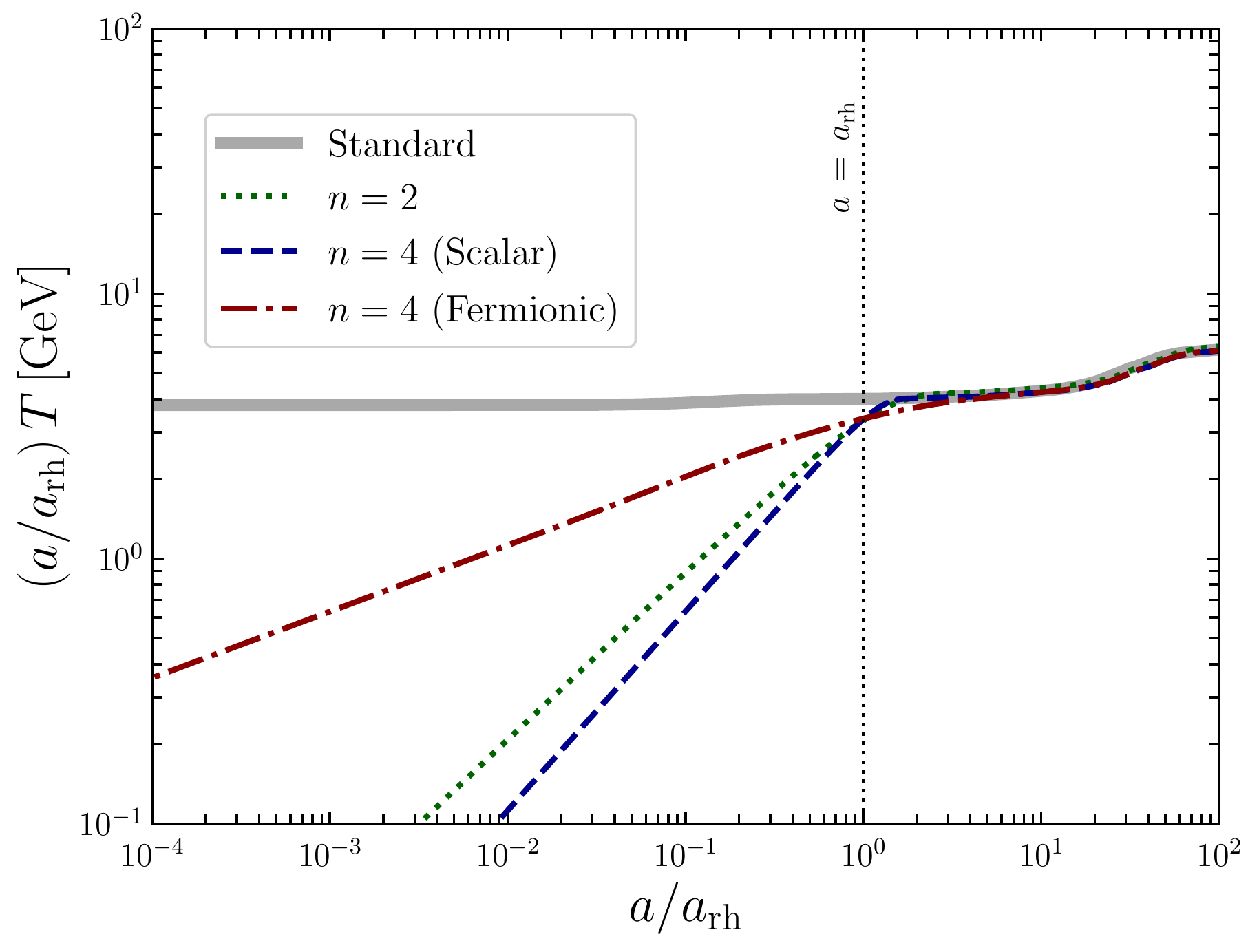}
    \includegraphics[width=\sepf\columnwidth]{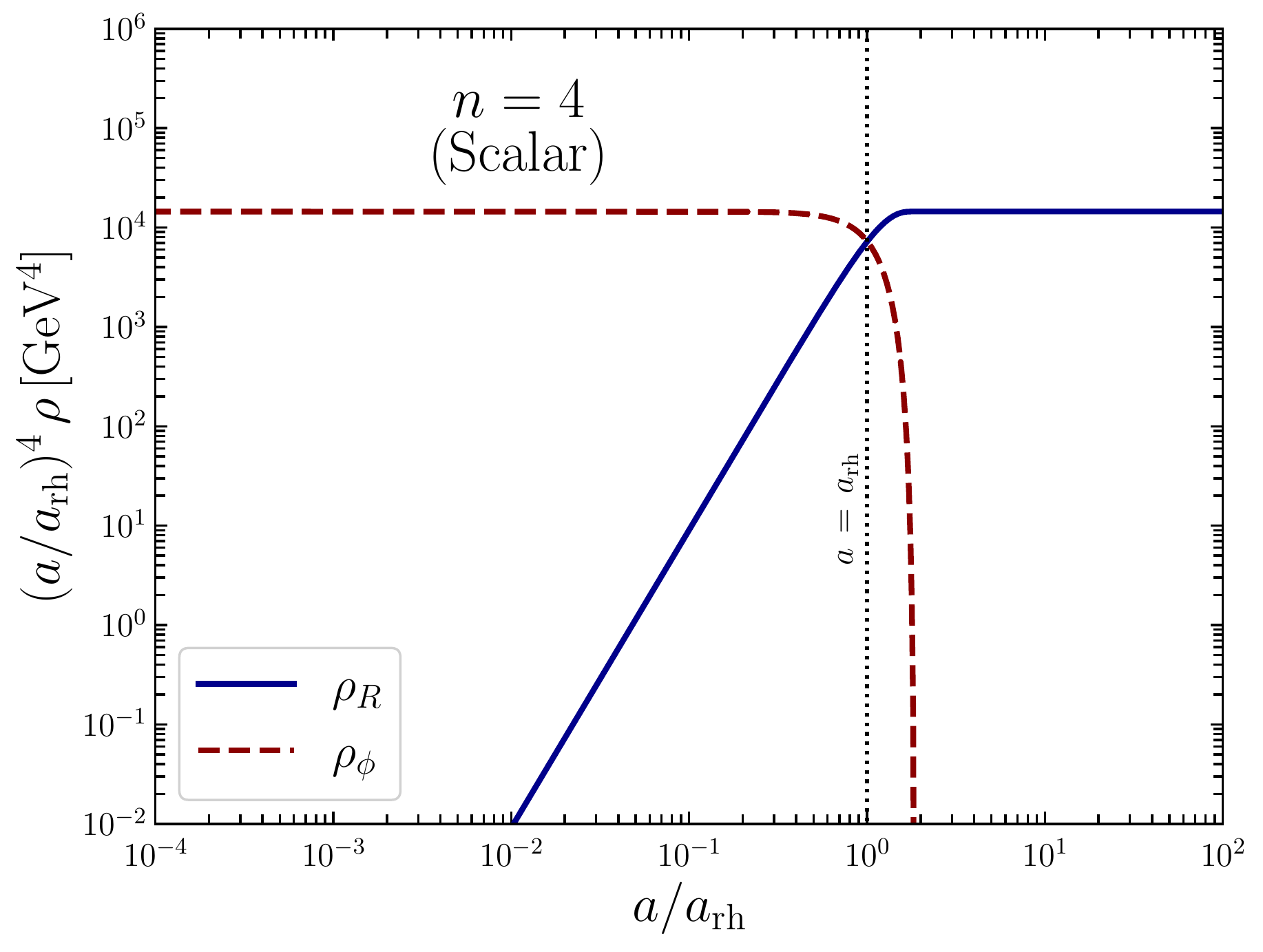}
    \includegraphics[width=\sepf\columnwidth]{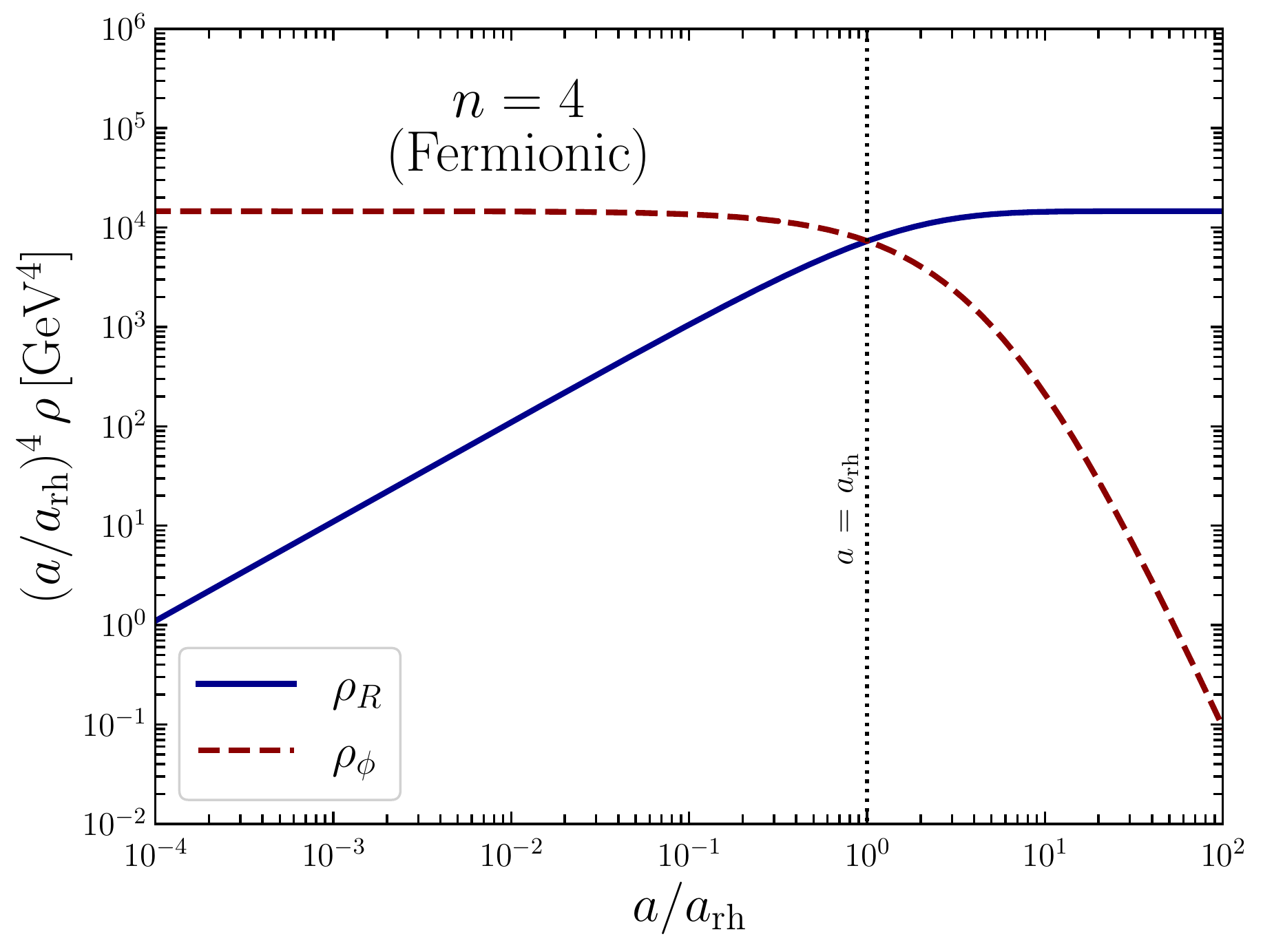}
    \caption{Evolution of the energy densities of the SM radiation $\rR$ and inflaton $\rp$ for the case $n = 2$ (top left), and $n = 4$ for a scalar (bottom left) or fermionic (bottom right) decay.
    The top right panel shows the evolution of the SM bath temperature $T$ as a function of the scale factor $a$ for the standard case (that is, a very high reheating temperature), and for the low reheating scenario with $\Trh = 4$~GeV for $n = 2$ and 4.
    The dotted vertical lines correspond to $a = \arh$.
    }
    \label{fig:d1}
\end{figure} 
The top left panel of Fig.~\ref{fig:d1} shows the evolution of the energy densities of the SM radiation (solid blue) and inflaton (dashed red), for the case $n = 2$ and taking $\Trh = 4$~GeV.
The curves are normalized with the factor $(a/\arh)^4$ so that the usual scaling after the end of reheating $\rR(a) \propto a^{-4}$ is clear. In this case, the inflaton oscillates in a quadratic potential, and therefore its mass and decay width are constant.
During reheating, $\rp(a) \propto a^{-3}$ and $\rR(a) \propto a^{-3/2}$, and therefore $T(a) \propto a^{-3/8}$, as shown in the top right panel.
As a reference, the top-right panel also shows the evolution of the bath temperature in the scenario of a high-reheating temperature. In that case, $T(a) \propto a^{-1}$, up to changes in the relativistic degrees of freedom $\gss$, cf. Eq.~\eqref{eq:TSM}.
After the end of reheating, the inflaton decays exponentially fast, and standard cosmology is recovered.

Alternatively, the bottom panels of Fig.~\ref{fig:d1} show the cases $n = 4$, where the inflaton instead oscillates in a quartic potential, and its energy density behaves like free radiation, $\rp(a) \propto a^{-4}$. In this case, the inflaton decay width is not constant in time. The panel on the left (right) describes the evolution when $\phi$ decays exclusively into scalars (fermions). During reheating, the scaling of $\rho_R$ for scalar decay is slower than for $n = 2$, evolving like $a^{-1}$ and leading to a steeper curve. In contrast, for fermionic decay, the energy density scales like $a^{-3}$, so the slope is milder.
The corresponding evolution of the SM temperature is also shown in the top right panel, showing the scaling $a^{-1/4}$ and $a^{-3/4}$, for the scalar and fermionic cases, respectively.

Having understood possible cosmological histories for the background, in the next section the evolution of the DM number density will be studied, taking particular care to the case where the bulk of the DM is produced during reheating.

\section{Dark Matter Genesis}
\label{sec:DM_density}
Now, we will focus on both thermal and non-thermal production for DM.
An important fact is that the evolution of the DM number density $n_s$ for the WIMP and FIMP mechanisms is governed by the same Boltzmann equation,
\begin{equation} \label{eq:boltzdm}
    \dfrac{dn_s}{dt} + 3\, H\, n_s = - \sv \left(n_s^2 - n_\text{eq}^2\right),
\end{equation}
where $\sv$ is the 2-to-2 thermally-averaged cross-section for the pair annihilation of DM particles into a couple of SM states and $n_\text{eq}$ corresponds to the equilibrium DM number density, for a Bose-Einstein statistics and a single degree of freedom, in the present case.\footnote{It is important to emphasize that we are assuming that DM is only produced from the scattering of SM particles. In particular, we disregard the possible direct production from inflaton decays. This is typically a good assumption as long as the branching fraction $\phi \to s\, s$ is smaller than $\sim 10^{-4}\, \mdm/(100~\text{GeV})$~\cite{Drees:2017iod, Arias:2019uol}.}

As the DM production during reheating will be studied, instead of using the usual DM yield $Y \equiv n_s/s_R$ where
\begin{equation}
    s_R(T) \equiv \frac{2 \pi^2}{45}\, \gss(T)\, T^3
\end{equation}
is the SM entropy density, it is convenient to rewrite Eq.~\eqref{eq:boltzdm} as a function of the comoving variable $N \equiv n_s \times a^3$ and $a$ as
\begin{equation}
    \frac{dN}{da} = - \frac{\sv}{a^4\, H} \left(N^2 - N_\text{eq}^2\right),
\end{equation}
where $N_\text{eq} \equiv n_\text{eq} \times a^3$.
It is solved using an \texttt{implicit Euler} method, from very early times to the present universe.
To match the entire observed DM relic density it is required that
\begin{equation}
    \mdm\, Y_0 = \frac{\Omega h^2\, \rho_c}{s_0\, h^2} \simeq 4.3 \times 10^{-10}~\text{GeV},
\end{equation}
where $Y_0$ is the asymptotic value of the DM yield at low temperatures, $s_0 \simeq 2.69 \times 10^3$~cm$^{-3}$ is the present entropy density~\cite{ParticleDataGroup:2020ssz}, $\rho_c \simeq 1.05 \times 10^{-5}~h^2$~GeV/cm$^3$ is the critical energy density of the universe, and $\Omega h^2 \simeq 0.12$ is the observed DM relic abundance~\cite{Planck:2018vyg}.\footnote{We note that other codes that solve the evolution of the DM density in the context of non-standard cosmologies exist (although they do not explore the reheating scenario), such as the ones presented in Refs.~\cite{Dutra:2021phm, Cheek:2022dbx, Cheek:2022mmy, Karamitros:2023uak}.}

\begin{figure}[t]
    \def\sepf{0.495}
    \centering
    \includegraphics[width=\sepf\columnwidth]{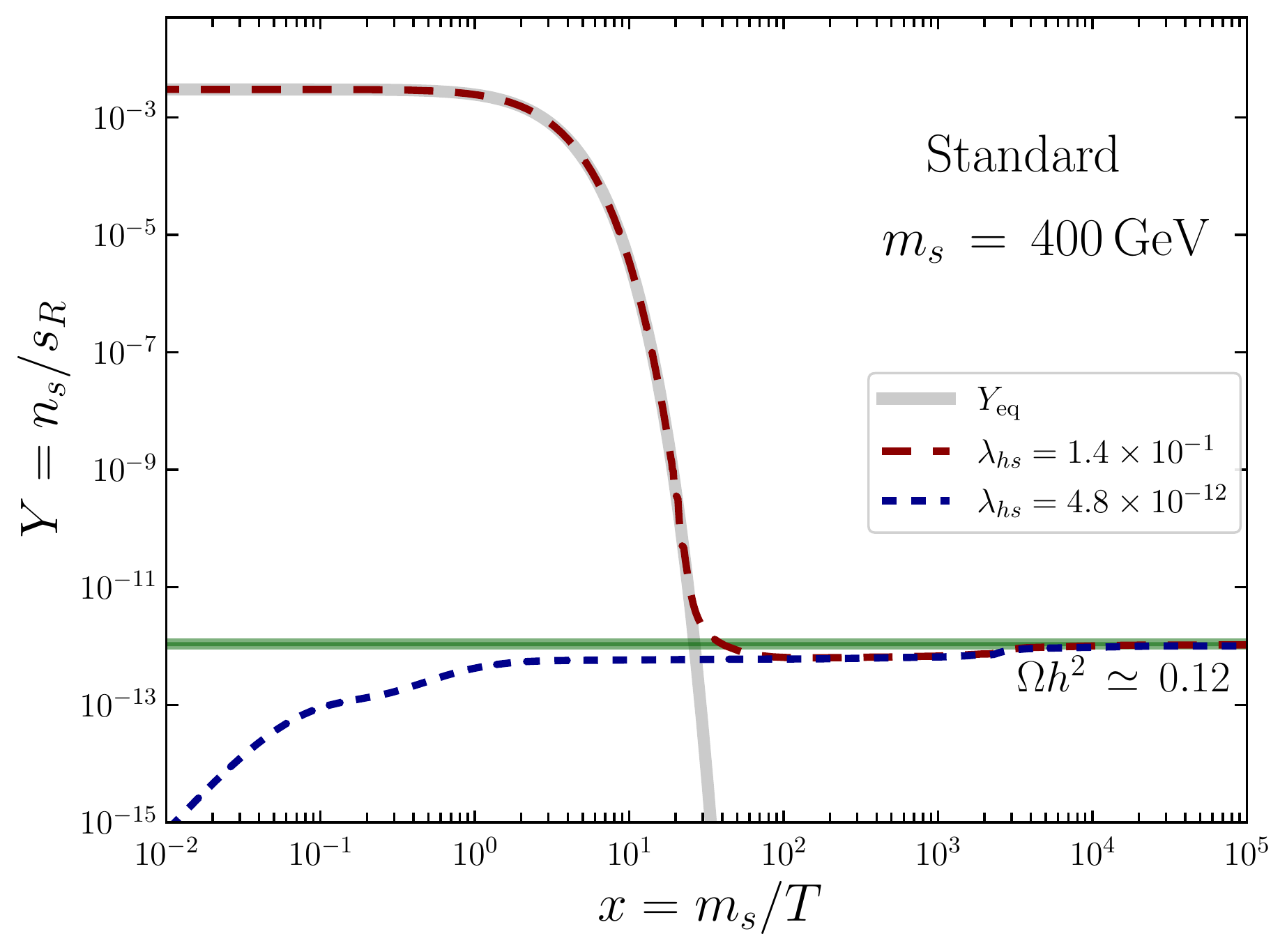}
    \includegraphics[width=\sepf\columnwidth]{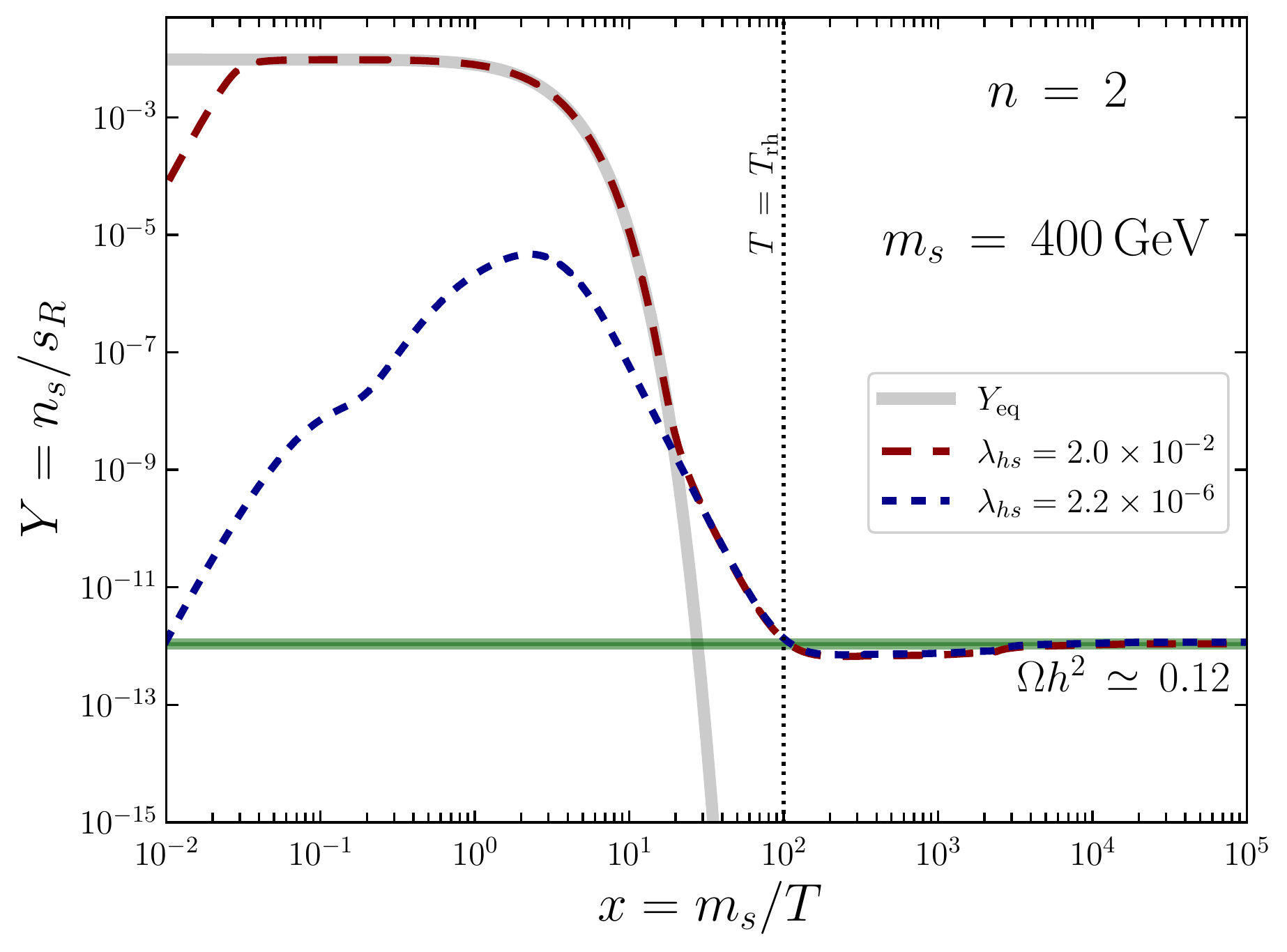}
    \includegraphics[width=\sepf\columnwidth]{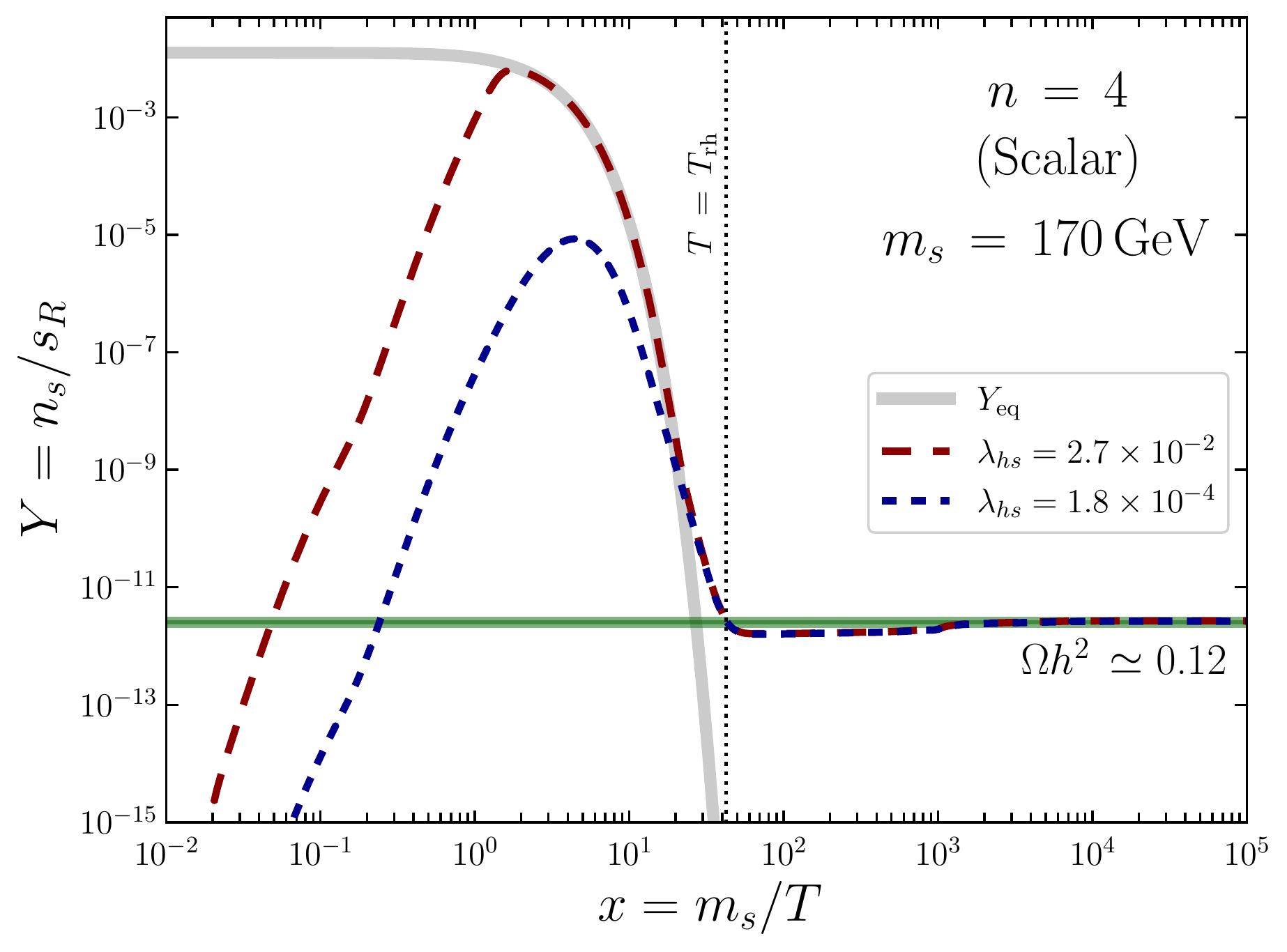}
    \includegraphics[width=\sepf\columnwidth]{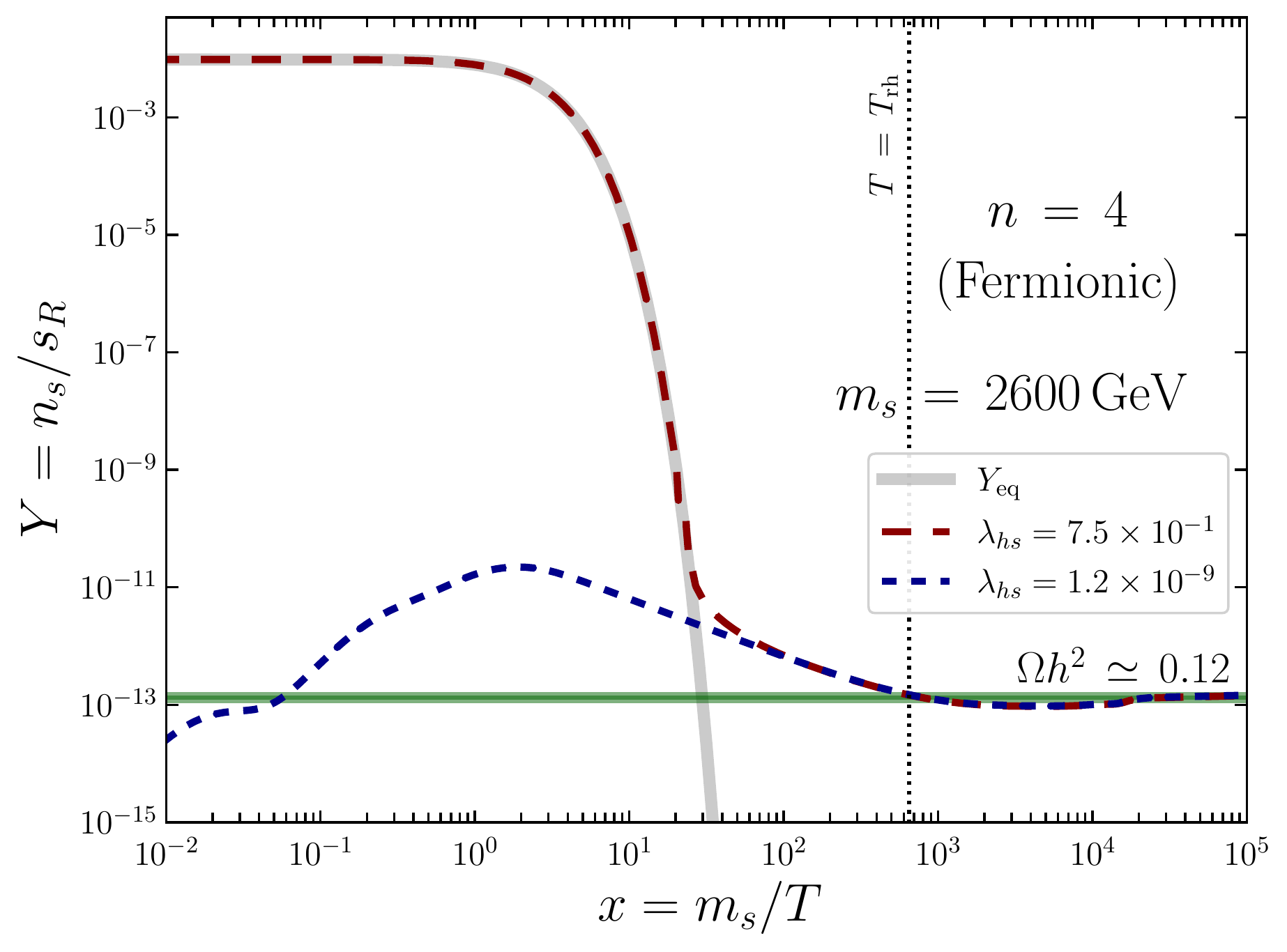}
    \caption{Evolution of the DM yield $Y$ as a function of the inverse of the SM temperature $T$ for different reheating scenarios.
    Top left: Standard scenario with high reheating temperature. Top right and bottom panels: $\Trh = 4$~GeV and $n = 2$ or $n = 4$, respectively.
    The gray bands correspond to the equilibrium yield, the thick green horizontal lines correspond to the bound $\Omega h^2 \simeq 0.12$, and the dotted black lines correspond to $T = \Trh$.
    Each plot presents two solutions that fit the whole observed DM abundance, one WIMP (red) and the other FIMP (blue).
    }
    \label{fig:d2}
\end{figure} 
The top left panel of Fig.~\ref{fig:d2} shows the evolution of the DM yield $Y$ as a function of the variable $x \equiv \mdm/T$ for the standard cosmological scenario where the reheating temperature is very high so that the bulk of the DM is produced in the radiation-dominated era.
The gray band corresponds to the equilibrium yield, while the thick green horizontal line corresponds to the bound $\Omega h^2 \simeq 0.12$.
For the DM mass $\mdm = 400$~GeV, two benchmark points that fit the entire observed DM abundance are presented.
The first case, for which $\lhs = 0.14$ (red dashed line), corresponds to the usual WIMP scenario, where the DM tracks the equilibrium density until it freezes out at a temperature $T \sim \mdm/25$.
The second one, corresponding to $\lhs = 4.8 \times 10^{-12}$ (blue dashed line), is an example of IR freeze-in production. Here, as the interaction rate is very suppressed because of the smallness of the portal coupling, DM never reaches chemical equilibrium, and the bulk of its production happens at $T \sim \mdm$.\footnote{The maximum of the DM production at $T \sim \mdm$ is typical from the IR-FIMP scenario with a light mediator, in this case, the Higgs boson $m_h \ll \mdm$.}

The other panels of Fig.~\ref{fig:d2} show low-reheating scenarios. As a general comment, we note that because most of the DM is produced during reheating, the yield $Y$ suffers from a high entropy dilution, meaning that a larger DM production is required to satisfy the observed relic density. On the one hand, for the WIMP case, the required overabundance is generated by having an earlier freeze-out. Interestingly, the presence of the inflaton leads to an increase of the Hubble expansion rate (see Eq.~\eqref{eq:H}), which forces the freeze-out to occur earlier. However, we find that if we do not modify the coupling, then the final relic density still comes out too small. In other words, the entropy dilution between the early freeze-out temperature and $\Trh$ is not compensated for by the larger production due to the change in Hubble rate. Thus, to successfully reproduce the correct relic density, the WIMP case requires smaller couplings, such that freeze-out occurs even earlier.

On the other hand, for the FIMP case, the final DM yield is proportional to the production rate and therefore to $\lhs^2$. The solution is straightforward; in order to counterbalance the injection of entropy, it is necessary to increase the coupling. It is worth noting that most of the production remains at $T \sim \mdm$; at higher temperatures $\mdm \lesssim T \lesssim \Trh$ the effect of entropy dilution clearly appears.

The aforementioned considerations can be seen in the upper right and lower panels of Fig.~\ref{fig:d2}, where the evolution of the DM yield is shown again, but this time for a low reheating temperature $\Trh = 4$~GeV. On the top right, we have the same mass as before, $\mdm = 400$~GeV, and take $n = 2$. Two solutions that match the abundance of the DM relic abundance are shown, $\lhs = 2 \times 10^{-2}$ (WIMP, red) and $\lhs = 2.2 \times 10^{-6}$ (FIMP, blue), following our expectations. For the WIMP case, it is interesting to note that, since we have lower values for the portal coupling $\lhs$, thermalization takes longer, occurring for this particular benchmark at $x \sim 3 \times 10^{-2}$.
Finally, a similar trend can be seen in the bottom panels of Fig.~\ref{fig:d2}, where the cases with $n = 4$ are shown for scalar (left) and fermionic (right) decays.
The stronger (weaker) scaling of the temperature for scalar (fermionic) decay, cf. Fig.~\ref{fig:d1}, delays (accelerates) chemical equilibrium. This is also reflected in the different evolutions of the DM yield after the DM abundance is settled.

\begin{figure}[t]
    \centering
    \includegraphics[width=0.9\columnwidth]{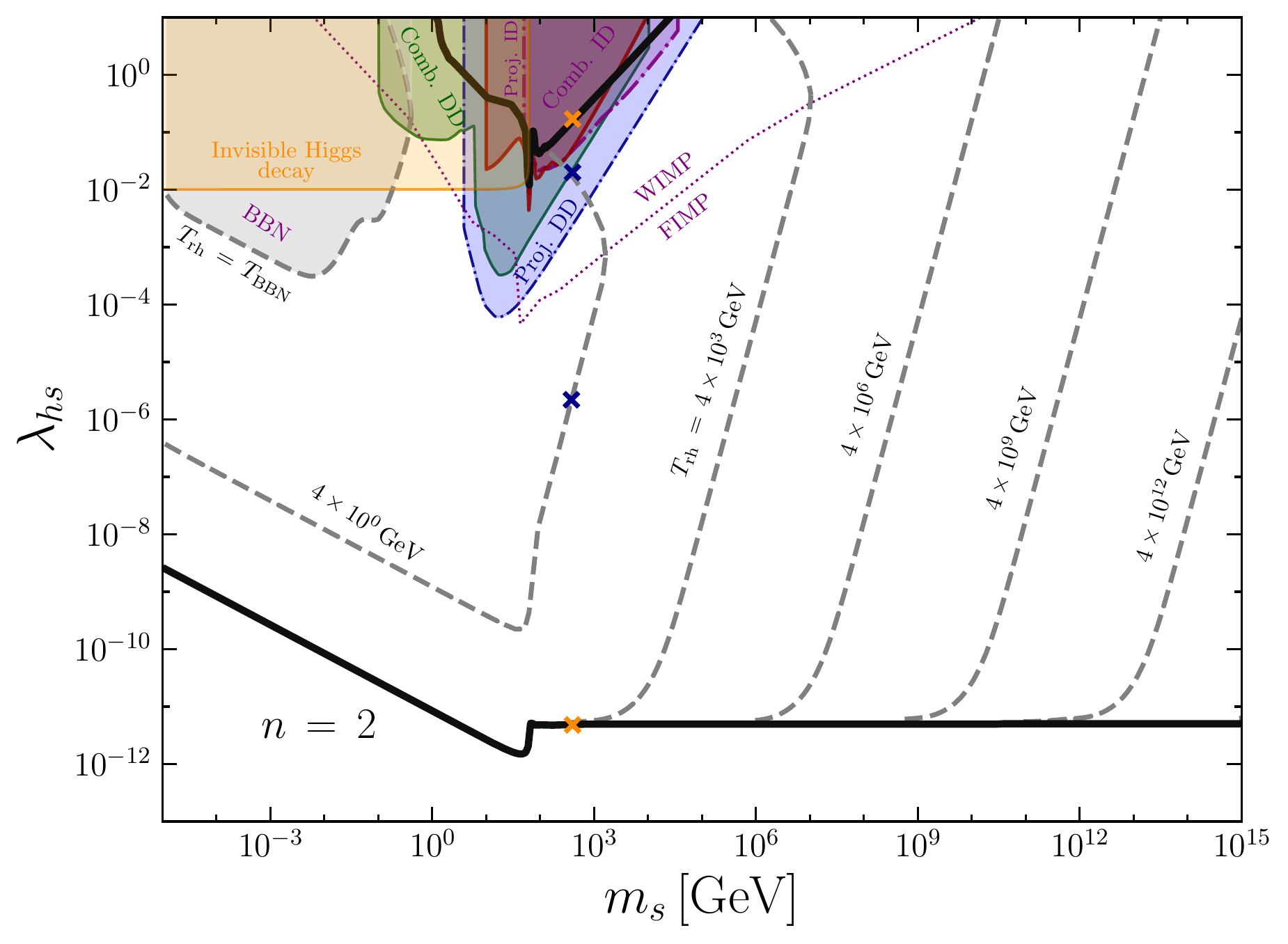}
    \caption{Parameter space reproducing the observed DM abundance, for an inflation oscillating in a quadratic potential ($n = 2$).
    The thick black lines correspond to the standard scenario with a very high reheating temperature, where DM is produced in the radiation-dominated epoch.
    The gray dashed lines correspond to lower reheating temperatures.
    The dotted purple line separates thermal (above) from non-thermal (below) DM production.
    The crosses show the benchmark points used in Fig.~\ref{fig:d2}.
    Shaded areas represent the experimental constraints and projections described in the text.
    }
    \label{fig:n2}
\end{figure} 
The same exercise of finding points that fit the observed value of the DM relic abundance can be systematically carried out by scanning over the DM mass $\mdm$ and the Higgs portal coupling $\lhs$, with results shown in Fig.~\ref{fig:n2}. Here, we restrict ourselves to perturbative couplings and to a mass range $10~{\rm keV}< \mdm < 10^{15}$~GeV; lower masses correspond to warm DM and are in tension with structure formation, while the upper bound on the inflationary scale $H_I^\text{CMB} \leq 2.0 \times 10^{-5}~M_P$~\cite{Chung:1998rq, Planck:2018jri} limits the production of heavier particles from the SM plasma~\cite{Bernal:2019mhf, Barman:2021ugy, Barman:2023ymn}.

Let us first focus on the thick black lines that show the required values for $\lhs$ as a function of $\mdm$ in the standard case with high reheating temperature. Two separate curves can be distinguished, identified as the WIMP and FIMP solutions for large and small values of $\lhs$, respectively. The WIMP solution for the SSDM model has interesting features, but for our purposes, only two need to be pointed out. First, a funnel appears at $\mdm \simeq m_h/2 \simeq 63$~GeV, which corresponds to the resonant production of a couple of DM particles through the $s$-channel exchange of a real Higgs boson. Second, perturbativity and unitarity conditions restrict the mass of DM to be between 1~GeV $\lesssim \mdm \lesssim 300$~TeV~\cite{Griest:1989wd, Smirnov:2019ngs}.
On the contrary, for the FIMP case, the dominant production channels correspond to $W^+W^- \to ss$ for $\mdm \gg m_h/2$ and to $h\to ss$ for $\mdm \ll m_h/2$. It is worth emphasizing that, as the FIMP paradigm is a non-thermal mechanism, it has an intrinsic dependence on the initial conditions, in particular, on other production mechanisms during inflation and reheating. Here, we follow the standard assumption and set the initial density of the DM to zero. In that sense, this line must be understood as an upper bound.

The impact of having a low reheating temperature on DM production is also shown in Fig.~\ref{fig:n2}, with gray dashed lines corresponding to different values of $\Trh$. We show solutions for reheating temperatures between $\Trh = T_\text{BBN}$ and $4 \times 10^{12}$~GeV. Several comments are in order. First, the presence of a low reheating temperature can only have an impact on the relic density (leading to the dashed gray lines diverging from the solid thick black lines) if the DM production occurs before reheating ends. For WIMPs, this means that $\Trh \lesssim \mdm/25$, while for FIMPs we need $\Trh \lesssim \mdm$ ($\Trh \lesssim m_h/2$), for $\mdm$ heavier (lighter) than $m_h/2$, respectively.

Second, as observed earlier, if DM is produced during reheating, the large injection of entropy has to be compensated for by higher production. This means that WIMPs require smaller couplings, whereas FIMPs require larger couplings. As a result, these two solutions can merge in the presence of a low reheating temperature, as can be seen, for example, in the cases with $\Trh = 4$~GeV and 4~TeV. This is in stark contrast to the standard cosmological scenario, where both solutions are completely disconnected.

In the figure, the purple dotted line shows the transition between the WIMP and the FIMP regimes. This was estimated analytically by comparing the expansion rate $H(T)$ with the interaction rate $\sv(T) \times n_\text{eq}(T)$.
Above that line, chemical equilibrium is reached and, therefore, DM is produced thermally via the WIMP mechanism.
Interestingly, this implies that WIMP becomes a viable scenario for higher masses of DM, well beyond the usual unitary limit $\mdm \lesssim 300$~TeV~\cite{Griest:1989wd, Smirnov:2019ngs, Bhatia:2020itt}.
Inversely, below the purple dotted line, chemical equilibrium is never reached, and hence DM was produced non-thermally.
Finally, the crosses shown in Fig.~\ref{fig:n2} correspond to the benchmark points used in Fig.~\ref{fig:d2}.

Let us now turn to the current constraints and future prospects, also shown in Fig.~\ref{fig:n2}.
First, the regions corresponding to $\Trh < T_\text{BBN}$ are ignored and shown in gray shaded.
Second, recent results from the ATLAS and CMS collaborations using 139~fb$^{-1}$ collisions at $\sqrt{s} = 13$~TeV bound the invisible Higgs boson branching ratio to Br$_\text{inv}\leq 0.11$ at $95\%$~CL~\cite{ATLAS:2022yvh, ATLAS:2023tkt, CMS:2023sdw}.
The Higgs decay into a couple of DM particles contributes to its invisible decay if the latter is light enough, with a partial width given by
\begin{equation}
    \Gamma_{h\to ss} = \dfrac{\lhs^2}{8\pi}\, \dfrac{v^2}{m_h} \sqrt{1 - \left(\dfrac{2\, \mdm}{m_h}\right)^2}\,.
\end{equation}
Taking into account the total Higgs decay width $\Gamma_h\simeq 4.07$~MeV~\cite{Djouadi:2005gi, ATLAS:2023tkt}, it follows that $\lhs \lesssim 5\times10^{-3}$ for $\mdm \ll m_h/2$, so that the orange region in the figure is ruled out.

The impact of direct and indirect detection experiments on our parameter space is estimated using \texttt{micrOMEGAs}~\cite{Belanger:2020gnr}. In the case of direct detection, we calculate the interaction cross-section for elastic scattering of DM off of nucleons for the spin-independent case. In the figure, the green region shows a simple combination of these experiments, labeled ``Comb. DD''. Here, we include XENON1T~\cite{XENON:2018voc}, LUX-ZEPLIN (LZ)~\cite{LZ:2022lsv}, CDMSlite~\cite{SuperCDMS:2015eex}, EDELWEISS~\cite{Lattaud:2022jnq} and XENONnT~\cite{XENON:2023cxc}. Furthermore, the blue regions (``Proj. DD'' label) show the projections of the XLZD consortium~\cite{Aalbers:2022dzr}.
From these regions, we find that another interesting feature arising from the WIMP-FIMP transition lies in the fact that the parts of the parameter space favored by the FIMP production are already excluded by XENONnT and LZ, and can be further explored by next-generation direct detection experiments such as XLZD.

Finally, in the case of indirect detection, we calculate the total annihilation cross-section. This is shown in the red region (``Comb. ID'' label), where we combine different limits of indirect detection of DM, such as MAGIC and Fermi-LAT observations~\cite{MAGIC:2016xys}, and H.E.S.S.~\cite{HESS:2022ygk}. We also included bounds from the analysis derived from CMB observations~\cite{Kawasaki:2021etm}, and antiproton AMS-02 data~\cite{Cui:2018klo}. We highlight the curves, in the purple region, (``Proj. ID'' label) for the prospect limits for CTA~\cite{CTAConsortium:2017dvg} and SWGO~\cite{Viana:2021smp}. 

\begin{figure}[t]
    \centering
    \def\sepf{0.9}
    \includegraphics[width=\sepf\columnwidth]{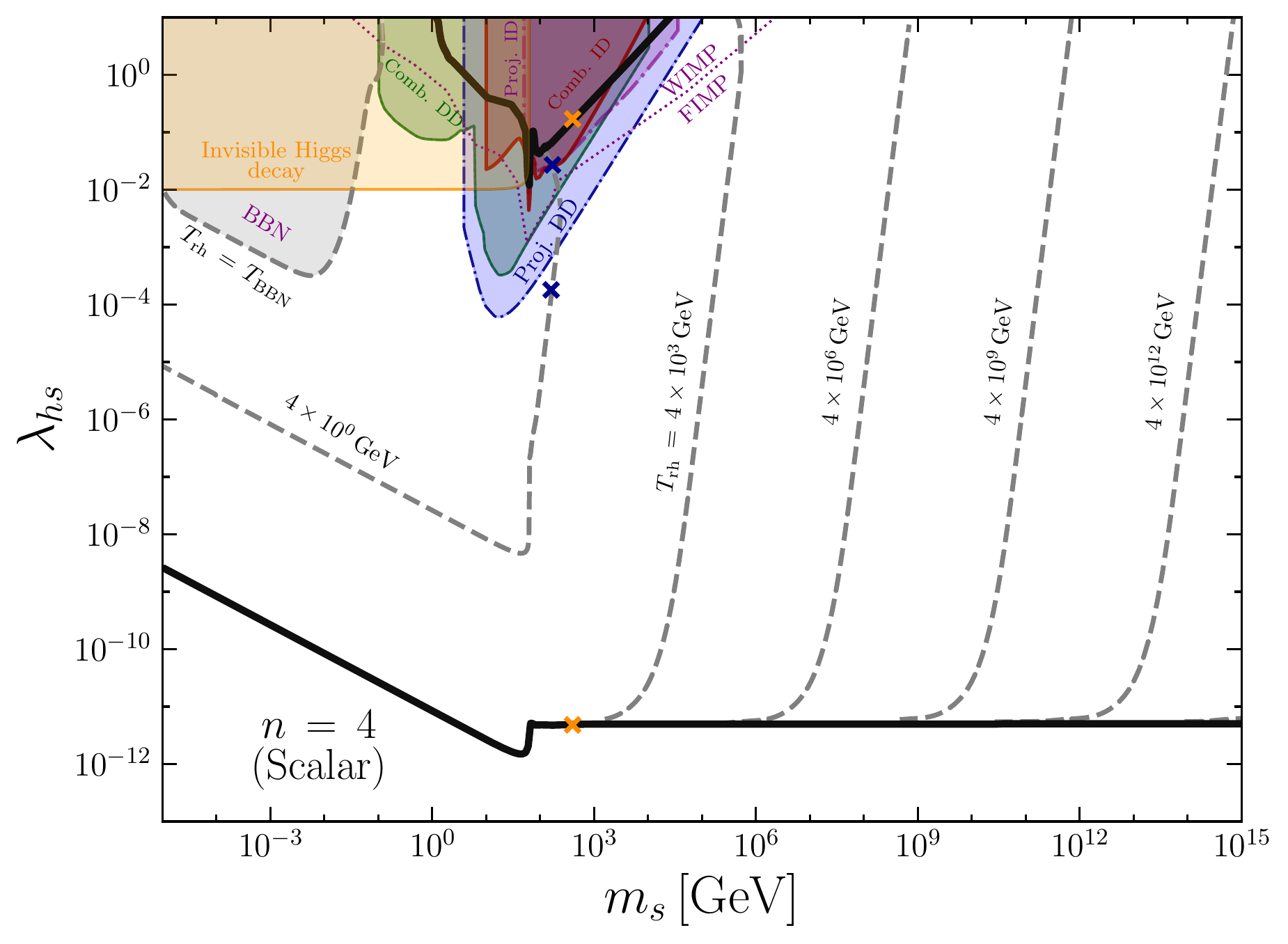}
    \includegraphics[width=\sepf\columnwidth]{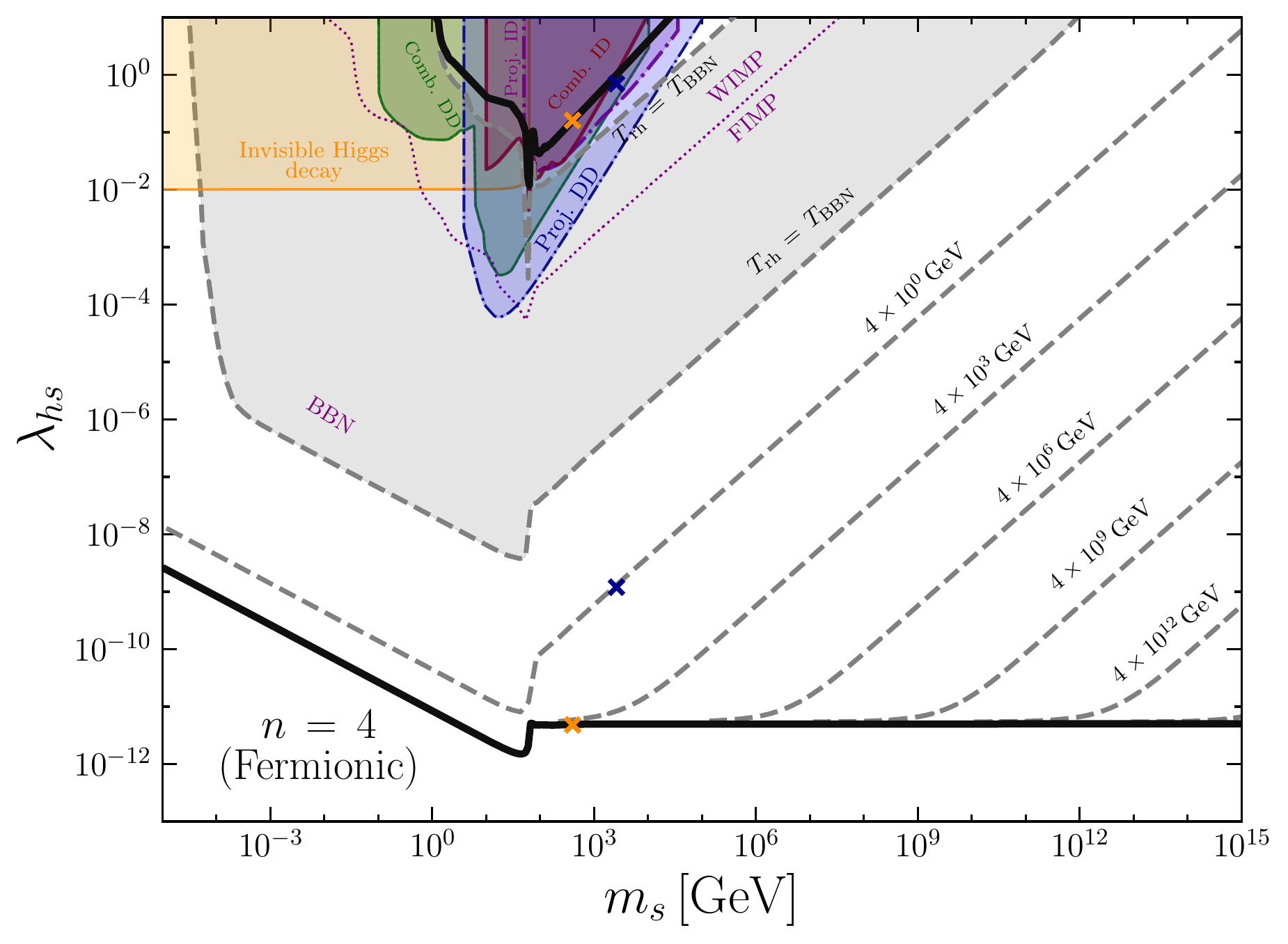}
    \caption{The same as in Fig.~\ref{fig:n2}, but with $n=4$ scalar decay (top panel) and $n=4$ fermionic decay (bottom panel).}
    \label{fig:n4}
\end{figure} 
Similarly to Fig.~\ref{fig:n2}, Fig.~\ref{fig:n4} shows the required values for the coupling $\lhs$ as a function of the DM mass $\mdm$, now for $n = 4$ (that is, an inflaton oscillating in a quartic potential) and a scalar (top) and a fermionic (bottom) reheating scenario.
For the WIMP region, the available parameter space shrinks with respect to the case $n = 2$, and consequently the region where DM is produced non-thermally increases. This is due to the fact that the dilution effect becomes less prominent with larger $n$, and therefore no large suppression of $\sv$ is required to compensate for this.
This effect is particularly visible in the fermionic case with $n = 4$, where there is a large range for masses spanning from $10^{-4}$~GeV to $10^{12}$~GeV where very small reheating temperatures, in conflict with BBN, are required.
Additionally, it is interesting to note that, contrary to the $n=2$ and scalar $n=4$ case, for fermionic decay with $n=4$ there is no smooth transition between the WIMP and FIMP solutions. 
This means that for a given $\Trh$, there are two disconnected solutions, one for WIMP and one for FIMP scenarios. In particular, this can be seen in the case $\Trh = T_\text{BBN}$ in the lower panel of Fig.~\ref{fig:n4}.
This feature comes from the fact that the temperature of the SM bath during reheating (for the fermionic case) tends to show a scaling similar to the free radiation for large values of $n$, being equal if $n > 7$~\cite{Bernal:2022wck}.

\section{Conclusions} \label{sec:concl}
Despite enormous experimental efforts over the last decades, the nature of dark matter (DM) remains one of the most complex and fundamental questions of particle physics.
Regarding its genesis in the early universe, it has been typically assumed that the DM is produced thermally or non-thermally as a weakly or feebly interacting massive particle (WIMP or FIMP, respectively).
Moreover, for the cosmological side, one usually works under the assumption of the so-called standard cosmology, in which the early universe energy density was dominated by standard model (SM) radiation from the end of inflationary reheating, until the BBN epoch. Additionally, the end of reheating is taken to happen at a very high energy scale, so that DM is produced when the universe is dominated by SM radiation.

In this work, we focus on an alternative possibility in which DM is produced during (and not after) inflationary reheating. In particular,  we investigate the production of DM during reheating in scenarios with a low reheating temperature, where the inflaton field $\phi$ oscillates around the minimum of a generic monomial potential $V(\phi) \propto \phi^n$, while decaying into fermionic or scalar states.
As a working example, we take the singlet-scalar DM (SSDM) model and explore the portal coupling so that DM could be a WIMP or a FIMP.
We implemented the model in \texttt{MadGraph} to obtain the relevant interaction rates and fully numerically solve the Boltzmann equation for the DM and the background.
The strong entropy injection caused by the inflaton decay has to be compensated by a reduction of the portal coupling in the case of WIMPs, or by an increase in the case of FIMPs.
Additionally, we pinpoint the smooth transition between the WIMP and FIMP regimes in the case of low reheating temperature.
Interestingly, while in the WIMP case, low-reheating temperatures favor smaller couplings to the SM particles, and therefore DM could easily evade experimental constraints, for the FIMP case larger couplings are needed, making this non-thermal production mechanism testable in present direct detection experiments like XENONnT and LZ. Next-generation DM direct detection experiments will further probe the viable parameter space of FIMP DM with low-reheating temperatures.
We emphasize that although these results correspond to the SSDM model, the conclusions presented here are expected to be very generic.

\acknowledgments
JS and JJP acknowledge funding from the {\it Dirección de Gestión de la Investigación} at PUCP, through the grant DGI-2021-C-0020.
NB received funding from the Spanish FEDER / MCIU-AEI under the grant FPA2017-84543-P.

\bibliographystyle{JHEP}
\bibliography{biblio}
\end{document}